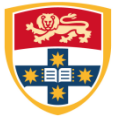

SCHOOL OF INFORMATION TECHNOLOGIES

# ON THE USABILITY OF HADOOP MAPREDUCE, APACHE SPARK & APACHE FLINK FOR DATA SCIENCE

## TECHNICAL REPORT 709

**BILAL AKIL, YING ZHOU, UWE RÖHM**

**MARCH 2018**

# Technical Report:
# On the Usability of Hadoop MapReduce, Apache Spark & Apache Flink for Data Science

Bilal Akil     Ying Zhou     Uwe Röhm

University of Sydney, Australia

`{bilal.akil,ying.zhou,uwe.roehm}@sydney.edu.au`

March 20, 2018

**Abstract**

Distributed data processing platforms for cloud computing are important tools for large-scale data analytics. Apache Hadoop MapReduce has become the de facto standard in this space, though its programming interface is relatively low-level, requiring many implementation steps even for simple analysis tasks. This has led to the development of advanced dataflow oriented platforms, most prominently Apache Spark and Apache Flink. Those platforms not only aim to improve performance through improved in-memory processing, but in particular provide built-in high-level data processing functionality, such as filtering and join operators, which should make data analysis tasks easier to develop than with plain Hadoop MapReduce. But is this indeed the case?

This paper compares three prominent distributed data processing platforms: Apache Hadoop MapReduce; Apache Spark; and Apache Flink, from a usability perspective. We report on the design, execution and results of a usability study with a cohort of masters students, who were learning and working with all three platforms in order to solve different use cases set in a data science context. Our findings show that Spark and Flink are preferred platforms over MapReduce. Among participants, there was no significant difference in perceived preference or development time between both Spark and Flink as platforms for batch-oriented big data analysis. This study starts an exploration of the factors that make big data platforms more – or less – effective for users in data science.

# Contents

# 1 Introduction

Across many scientific disciplines, automated scientific experiments have facilitated the gathering of unprecedented volumes of data, well into the terabyte and petabyte scale [18]. Big data analytics is becoming an important tool in these disciplines, and consequently more and more non-computer scientists require access to scalable distributed computing platforms. However, distributed data processing is a difficult task requiring specialised knowledge.

Distributed computing platforms were created to abstract away distribution challenges. One of the most popular systems is Apache Hadoop which provides a distributed file system, resource negotiator, scalable programming environment (namely MapReduce), and other features to enable or simplify distributed computing [3, 10]. While a large step in the right direction, effective use of this environment still requires familiarity with the functional programming paradigm and with a relatively low-level programming interface.

Following the success of Hadoop MapReduce, several newer systems were created introducing higher levels of abstraction. While MapReduce addresses the main challenges of parallelising distributed computations – including high scalability, built-in redundancy, and fail safety – newer systems including Apache Flink [2, 9] and Apache Spark [4, 30] focus more on the needs of efficient distributed data processing: dataflow control (including support for iterative processing), efficient data caching, and declarative data processing operators.

Scientists have now a choice between several distributed computing platforms, and to guide their decision several comparison studies have been published recently [6, 22, 23]. The primary focus of those studies was performance, which is perhaps not the primary problem for platforms built from the ground up with scalability in mind. More interesting is the question of the usability and ease-of-use of those platforms, given that they will be used by non-computer scientists. But there are no large-scale usability studies so far.

In this paper, we present the design, execution and results of a large usability study of three popular cloud computing platforms – Apache Hadoop MapReduce; Apache Spark; and Apache Flink – which was conducted as part of a cloud computing course for masters students from various backgrounds, including IT and data science. Participants of the study had to implement three different data analysis tasks with use cases from immunology and genomics. The first task was implemented using MapReduce, while the last two tasks were implemented in a crossed A/B test with half the class first using Flink and the other half Spark. The method used for usability study data analysis will also be discussed, wherein data was analysed using a Python Jupyter Notebook.

To the best of our knowledge, this is the largest usability study of modern data processing platforms. Our aim is to provide direction for selecting a suitable system considering factors other than solely performance. We also believe that our learnings will prove useful in guiding future in-class usability studies, and so discuss the successes and challenges met throughout our experience.

## 2 Related Work

There have been several comparison studies of distributed computing engines in the context of scientific applications before, which however typically focus on performance and scalability of the systems, somewhat neglecting usability metrics. For example, Bertoni et al. are comparing the same systems (Apache Flink and Apache Spark) with regard to genomics applications [6], but only report on differences in implementation techniques and runtime performance. Similar performance comparison studies of Spark and Flink with varying analytical workloads have been done by Marcu et al. [22] and by Perera et al. [25].

A comparison study more closely related to this work is by Mehta et al., who compare five big data processing systems (Apache Spark, SciDB, Myria, Dask and TensorFlow) with regard to their suitability and performance for scientific image analysis workflows [23]. This paper also gives a brief qualitative assessment of each system, however based on measuring lines of code and observing implementation issues [22]. A second study which also compares Apache Hadoop MapReduce, Spark and Flink in areas other that performance, such as usability, understandability and practicality [14], was performed but was based on the experience and views of its sole researcher instead of a cohort in a usability study.

We found no usability studies comparing the distributed systems in this paper. The usability study by Nanz et al. [24] compared concurrent programming languages, and while similar in how it subjected a university class to two different programming languages and compared the results, it had spanned only four hours and was set in a more controlled environment, and thus wouldn't face many of the challenges that our semester-long study would. The usability study by Hochstein et al. [19] compared the programming effort of two parallel programming models, and was also of a similar comparative nature, participant base, and a comparable time-frame, but heavily utilised instrumented compilers, which in our case would be impractical considering time restraints and system complexity. It also was focused on comparing effort in the form of development time and correctness, which we felt would not be sufficient to describe and compare the broader usability of a system.

## 3  Systems

Apache Hadoop MapReduce [10] has long been the de facto standard for large-scale data analytics, being one of the earliest systems available to abstract the challenges of distributed computing and fault tolerance, significantly reducing the barrier to entry that was present in the big data space.

Its success later led to the creation of systems which provided higher level approaches to distributed computing. Apache Spark [30] and Apache Flink (formerly Stratosphere) [9] are two prominent examples of such systems. The two are seen as common rivals, and have had much attention paid to their performance merits and pitfalls [22].

The focus of this study is instead on their usability. All three systems, in this context, will run using Apache Hadoop YARN [28] for resource management and HDFS [26] as the distributed file system. The following versions were used in the usability study: Apache Hadoop MapReduce v2.7.2; Apache Spark v2.1.1; Apache Flink v1.2.1. These were all the stable or highest non-beta versions at the time of preparation.

The high-level design of the systems, more from a usage than architecture perspective, will be described in this section.

## 3.1 Apache Hadoop MapReduce

As the name suggests, Apache Hadoop MapReduce is executed in the Hadoop ecosystem, typically utilising YARN for cluster management and HDFS as a distributed file system. Hadoop MapReduce facilitates the fault-tolerant, distributed execution of 'jobs', which encompasses the following processing steps:

1. Read input from HDFS blocks and split to mappers.
2. Map, applying a user-defined function (UDF).
3. No reducer: output one file per mapper and finish.
4. Optionally combine output from mappers using a UDF.
5. Partition, shuffle, sort and merge data into reducers. Default partition and sort behaviour can be overridden.
6. Reduce using a UDF.
7. Output one file per reducer to HDFS.

In Hadoop MapReduce, the driver is a Java class utilising the MapReduce Java package which configures, starts and waits for jobs to execute. It can access written data between jobs by reading their output, for instance from HDFS.

Alternatively, in streaming mode, the driver is instead a set of shell commands, where scripts are specified to act as the mapper, combiner and reducer, each operating via standard input and output. This technically allows any method of programming available throughout the cluster to be used, and may present other contextual advantages or disadvantages [11]. Chaining jobs would then become a matter of chaining shell commands.

The mapper and reducer are classes or scripts that operate on key value pairs. A mapper receives an iterator of key value pairs and can output zero or more key value pairs. A reducer receives one key and an iterator of values, or an iterator of key value pairs in sorted key order in Hadoop Streaming, and can output zero or more key value pairs. A combiner is a reducer that is executed on each mapper following mapping but prior to data being shuffled over the network.

Other distributed computing operations are implemented in terms of mapping and reducing. For instance, filter would be in the mapper, while joining and aggregation would be in one or both of a mapper and reducer, presenting different trade-offs [7]. Iteration can be implemented using a loop in the driver, and in that loop configuring and starting new jobs that use the previous completed jobs' output. Higher level systems have been created to improve support for or simplify iteration in MapReduce, such as Twister [12].

Listing 1: Apache Spark Python word count example as shown at: https://spark.apache.org/examples.html

```
text_file = sc.textFile("hdfs://...")
counts = text_file.flatMap(lambda line: line.split(" ")) \
             .map(lambda word: (word, 1)) \
             .reduceByKey(lambda a, b: a + b)
counts.saveAsTextFile("hdfs://...")
```

## 3.2 Apache Spark

Apache Spark turns input data into Resilient Distributed data sets (RDDs) which "lets programmers perform in-memory computations on large clusters in a fault-tolerant manner" [29]. Lazy transformations are applied to RDDs, creating new RDDs, where execution of said transformations do not occur until necessary for consumption by an 'action' – for instance for collection onto the driver or for storage into HDFS.

Spark's core API features various generic transformations and actions, and is fulfilled via one of three resource managers: Spark Standalone; Hadoop YARN; or Apache Mesos. Additional APIs have been built atop the core API to provide higher level support for various contexts. These APIs are provided for different programming languages. The core API currently supports Scala, Java, Python and R.

The driver is any program which utilises the core API and optionally the other more specialised APIs. It creates RDDs from various input sources, including the local file system or HDFS, and applies lazy transformations and actions to those RDDs. The driver can also be an interpreter, which is often useful for exploration or debugging.

Iteration can be performed similarly to Apache Hadoop MapReduce; using a loop in the driver. However, instead of configuring, starting and blocking on new jobs which write to and from HDFS, Spark would simply apply additional lazy transformations and move on looking for the next action.

Spark mostly performs in-memory computation in an attempt to minimise disk communication. This has the potential to provide speed improvements in various circumstances, including iteration, but can also degrade performance if memory is insufficient [15]. More work is being put towards improving memory management to improve resiliency and performance [22].

The core API operates on key value pairs or arbitrary objects. It includes transformations such as: `map`, `filter`, `reduceByKey`, `distinct`, `union`, `intersection`, `sortByKey`, `aggregateByKey`, `join`, and so forth. Actions include `saveAsTextFile`, `collect`, `count`, `countByKey`, `first`, `foreach`, `takeSample`, and so forth. Some transformations or operations can only operate on key value pairs.

With thanks to its high-level APIs, Apache Spark programs can end up looking quite simple, like the word count example in Listing 1. However, in reality users will need to understand various system internals, such as when data is shuffled, to support the design of efficient and scalable programs.

Fault tolerant, distributed stream processing is achieved in Apache Spark by using the Spark Streaming extension of the core API. It works by dividing or 'micro-batching' live input data streams into a 'discretized stream' or `DStream`, which is a sequence of RDDs [31], that can be operated on by the core API and additional streaming operations such as `window`. Other Spark APIs, including MLlib and GraphX, are also able to operate on `DStream`s. Apache Spark's method of micro-batching has been found to be slower but more resilient to failure, than native streaming in Apache Storm and Apache Flink [21].

## 3.3 Apache Flink

Apache Flink has changed much since its Stratosphere days. Thus, the information here is based on the Apache Flink v1.2 documentation found at `https://flink.apache.org`.

Flink is natively a stream processor where batch processing is represented as a special case of steaming – more specifically, bounded streaming with some adjustments to features such as fault tolerance and iteration. In Flink, users specify lazy streams and transformations which the engine then maps to a streaming dataflow using a cost-based optimiser. This dataflow is a directed acyclic graph (DAG) from sources to sinks, with transformation operators in between. Sinks trigger the execution of necessary lazy transformations.

The engine can similarly be run in standalone, Hadoop YARN, or Apache Mesos cluster modes. It provides APIs with different levels of abstraction. The lowest level API offers building blocks for stateful stream processing. The core `DataSet` (batch) and `DataStream` APIs sit atop that and are the most commonly used, with the table and SQL APIs sitting at atop them. Other libraries are provided to directly support various specific contexts. The core APIs support Java and Scala, with the data set API additionally supporting Python.

Similar to Apache Spark, the driver is any program which utilises the Flink APIs. It creates `DataSet`s or `DataStream`s from various input sources and applies lazy transformations to them, creating new `DataSet`s or `DataStream`s, until eventually directing them to a data sink.

Iteration can be achieved either using a loop in the driver, or via the `IterativeStream` or `IterativeDataSet` classes. The former is technically not iteration, but rather the driver looping and extending the DAG as necessary, which is limited in its scalability. The latter can be thought of as a single node in the DAG which performs a set of transformations iteratively, either using the last computed value or a solution set state that can be modified in each iteration.

Flink also primarily utilises in-memory computation to minimise disk communication. For robustness it implements its own memory management within the JVM, attempting to prevent out of memory errors by spilling to disk, reduce garbage collection pressure, and more.

The system does not operate on key value pairs, but requires 'virtual' keys for some operators like grouping. It handles arbitrary data types and provides additional support for tuples and objects by simplifying keying, for instance based on a tuple index or object property. Its core API supports a set of transformations that is largely similar to those in Apache Spark's core API. As a result, Apache Flink programs can also appear quite simple upon completion, but its users also will need to understand various system internals, such as when data is shuffled, to support the design of efficient and scalable programs.

## 4 Usability Study

The aim of this study is to compare the usability of three popular distributed computing systems: Apache Hadoop MapReduce; Apache Spark; and Apache Flink. The participants of the study are masters students from a cloud computing class at the University of Sydney, where the mentioned systems are taught. The focus is on data processing in the cloud, assessed with practical programming assignments. Stream processing is not covered in this usability study – all exercises are in the form of batch processing. As highlighted in the related work section, our study is quite novel and unique as the two closest existing usability studies of similar circumstance still differed fundamentally in scope and study duration.

We adapted effective study design considerations from those and other papers where possible, and besides applied our knowledge and best judgment in designing this usability study. This section will describe the background of the study, the study design and decisions that were made in its regard, and the strengths and challenges in its execution.

### 4.1 Background

The study was conducted as part of a regular master's level class on cloud computing at the University of Sydney. This class attracts a diverse student cohort because it is available for selection in several different degrees, most prominently including students studying computer science at either master's or undergraduate (4th year) level, or studying a Master of Data Science – which does not require a computer science background. Participation in this study was voluntary, so it was paramount to design and organise the usability study in such a way that students who did not opt-in to participate would not be at an advantage or disadvantage.

The class of 2017 was scheduled to start in early March, and consideration and preparation for this usability study began in early-mid February. Topics relevant to the usability study would begin being taught in early April through to late June, with the earlier weeks dedicated to general concepts like the cloud and datacentres. Thus the time-frame for the usability study would be 2.5 months, with 1.5 months preparation.

The of two authors of this study both held roles in execution of the unit of study while the study was being performed. A. Prof. Uwe Röhm was the unit of study coordinator and lecturer. Bilal Akil was the unit of study TA and one of its five tutors. Thus, once equipped with ethics approval, we had the means to reshape the class to better fit the usability study, provided that changes did not sacrifice students' quality of learning or learning outcomes, irrespective of one's decision to participate in the study. This responsibility was paramount and considered throughout design and execution of the study.

In previous years, this course covered Apache Hadoop MapReduce and Apache Spark, and this year a third system – Apache Flink – was taught too. Teaching material and exercises were prepared for all three systems and updated where necessary. Because of the diversity of the student cohort, this course supports both Java and Python as programming languages, which individual students can select to use as they prefer. This means six variants of exercise and assignment solutions (3 systems × 2 programming languages) were prepared.

The preparation time would not be long enough to address everything before the usability study commenced. Assignments and learning materials would still be developed and the cluster would still be updated as the usability study was executed. Time being a scarce resource was a factor that had to be considered in the study's design.

The assessment component of the class comprised three practical programming assignments that students would work on in pairs, plus a written final exam that is not part of this study. Students were provided with between three and four weeks to complete assignments, each of which were an increasingly complex series of distributed computing tasks in some domain.

## 4.2 Study Design

To be fair to all the class' students, we decided that they would all learn and use each of the three distributed computing engines, as opposed to dividing usage among them. This choice was made to avoid circumstances such as: "Why did (s)he use System X but I had to use System Y?"

Each assignment was targeted at different distributed computing engines (cf. Table 1), and provided students with an experience which they could then reflect upon to consider the usability of each system. The first assignment covered the lower level data cloud computing framework, Apache Hadoop MapReduce. This would place participants on a more equal starting point for comparison to the more modern, higher level data processing frameworks to come. This decision also complemented the existing course structure, where the fundamental and lower level distributed computing concepts are taught first. We suspected that the majority of participants would *not* prefer to use Hadoop MapReduce compared to the other systems (and you can see in Section 5 that this was indeed the case), and thus were not concerned about slight biases in favour of it that may be introduced by having used it first.

On the other hand, we were concerned that the order of usage for the next two systems could have a strong effect on their comparison results. To account for this, we decided to employ a crossed A/B test for the remaining two assignments: half of the participants would use Apache Spark for assignment 2 and Apache Flink for assignment 3, and the other half would do the opposite.

|              | System        | Scenario     | Data set                            | Tasks                                                        |
|--------------|---------------|--------------|-------------------------------------|--------------------------------------------------------------|
| **Assignment 1** | Hadoop MR     | Social media | Flickr: Photos, locations, and tags | Filter, transpose, join, aggregation, ranking                |
| **Assignment 2** | Flink \| Spark | Immunology   | Cytometry + experiment data         | Filter, transpose, join, aggregation, $k$-means clustering [16] |
| **Assignment 3** | Flink \| Spark | Genomics     | DNA microarray + patient data       | Filter, transpose, join, aggregation, Apriori algorithm [1]  |

Table 1: Overview on programming tasks used in study.

Therefore teaching and learning resources for both systems were made available at the roughly same time and depth.

### 4.2.1 Assignment Tasks

An overview of the assignment scenarios and tasks that were used in the usability study can be seen in Table 1. The main design considerations for the practical assignments were:

- The change from two systems and assignments in the past to three would be expected to increase the difficulty of the course. To empathise the assignments would be made smaller, requiring about two weeks for a pair to complete instead of three or four.

- Each assignment used a different data set to avoid having participants become accustomed to the same one.

- All data sets had schemas of similar complexity, with two to three tables given as CSV files that could be joined on a foreign key relationship, and one list-valued attribute that had to be transposed during querying.

- The first assignment required participants to exercise various distributed computing operations: map, reduce, filter, group and ranking. Being in Apache Hadoop MapReduce, the latter 3 required non-trivial implementation.

- The second and third assignments also covered declarative analysis with filtering, join and aggregation to allow comparison back with Hadoop MapReduce.

- Additionally, the last two assignments involved a task focused around some iterative data mining algorithm.

It was recognised that the difficulty of each assignment was variable, considering the changing systems (particularly from the lower level MapReduce in assignment 1 to the higher level systems), scenarios, data sets, and algorithms. However, these were all necessary either for the reduction of bias towards any particular system, or for the general flow of the course. Effects that assignment difficulty may have had on perceived usability has been explored in Section 5.

We intended on including assignment marks in the usability study data. As the unit of study had multiple tutors to teach classes and mark assignments, it was necessary (more than normal) to implement some form of marking that would reduce judgement and thus potential bias from individual markers. We took on the approach described in section 6.4 of the study by Nanz et al. [24], attempting to identify a set of mistakes and their weightings and marking backwards, so to speak. However, here it was necessary to release a marking rubric for students to access prior to completion of the assignment, and thus the set of potential mistakes had to be compiled beforehand, as opposed to after scanning submissions and categorising the mistakes that actually were made.

### 4.2.2 Data Analysis Scenarios

Each assignment was set in a different scenario to avoid any potential bias due to familiarity with a data set (cf. Table 1). As our aim is to study the usability of the distributed data processing platforms for non-computer scientists, we chose data analysis scenarios from social media, bioinformatics, and genomics.

Assignment 1 involved data analysis of Flickr data. The data set is an excerpt of real-world Flickr data including a hierarchical location attribute and multiple tags per photo given as a multivalued attribute. Students were asked to implement different analytical queries in MapReduce to identify: the number of photos taken at a certain locality level; the top 50 localities by number of photos; and the top 10 tags used for those photos.

Assignment 2 considered a scenario from immunology, involving real cytometry data from a study of infections with the West Nile Virus. We also provided fabricated metadata about the experiments including a multivalued attribute on the authors of the measurements. This assignment had two subtasks. Firstly, students had to determine the number of valid measurements per researcher, involving: filter; transpose; join; and aggregation operations, similar to assignment 1. Secondly, students had to complete a clustering task to identify similar cell measurements with regard to some given 3-dimensional cell markers using the $k$-means clustering algorithm [16].

Assignment 3 presents a genomics scenario requiring the analysis of a DNA microarray data set and patient metadata. This data set was synthetic – generated using the schema and data generator from the GenBase benchmark [27]. We modified the GenBase schema to allow the application of multiple disease codes per patient (instead of a single one) via a multivalued attribute. Students were first asked to find the number of cancer patients with certain active genes, covering the filter, transpose, join, and aggregation operations as in the previous two assignments. They further had to mine the microarray data in search of frequent combinations of expressed genes for certain cancer types using the Apriori algorithm [1].

### 4.2.3 Self-Reflection Surveys

We included a short survey as part of the assignment submission process, acting as a method of self-reflection for students after completing the assignment, and also as a primary source of the usability study's data. The survey was only available for completion following the submission of source code, thus captur-ing their view of each system upon completion of the relevant assignment, as opposed to say halfway through.

Although the assignments were completed in pairs, and thus only one source code submission was necessary for a pair of students, we emphasized that the self-reflection surveys were to be completed individually.

The surveys for each assignment included:

- A simple and standard usability survey: the System Usability Scale survey [8], as discussed in Section 4.2.4.

- A question directly asking which is their preferred system. This question does not apply to assignment 1.

- A question asking approximately how much time was spent working on the assignment. Options were separated into seven 4 hour bins, from 0-4 hours, to 20-24 hours and then 24+ hours.

- A text area to provide any textual feedback.

The first survey also included four questions to gauge students' prior programming experience. It asked how many years of programming experience the students considered themselves to have, from 0 to 9 and then 10+, and three Likert scale questions asking students for their perceived proficiency with Java, Python and shell environments, from "No proficiency" to "Very high proficiency".

We required all students to complete the survey as part of their learning outcomes, as we considered it a form of self-reflection. However, all questions were optional, allowing students to omit any if uncomfortable. Only participating students' survey data was included for the purpose of the usability study. We made sure to keep the survey brief and simple to reduce load on students and, perhaps optimistically, reduce fake responses and increase usability study participation.

### 4.2.4 System Usability Scale

The SUS [8] was used to provide a score for the usability of each system to be used for comparison. While programming frameworks or distributed systems may not be the intended application of this survey, we were unable to find anything suitable and more specifically targeted to this use case, and found that the SUS was general enough to be applied nonetheless, considering that our survey was explicitly focused on usability. In this regard we were trying something apparently new, and thus could not be sure of the applicability of the results.

The SUS is comprised of ten Likert scale questions. Odd numbered questions are positively natured, while even numbered are negative. The final score is compiled adding or subtracting those scores and applying a multiplier, becoming a single score between 0 and 100. Note that SUS scores are interval level – not ratio – meaning a SUS score of 80 is not necessarily twice as good as 40.

We decided to use this survey as it presented a good balance of ease and generality. Ease here refers to the survey only having ten questions of a simple nature. Generality is in how the survey only presents rather general statements about the 'system', making it applicable to a variety of systems even with significant differences between them, and avoiding references which wouldn't apply or would be confusing in a programming context, such as 'scenario' in the ASQ [20], or 'information' and 'interface' in the CSUQ [20].

Despite the availability of more succinct questionnaires, we found the SUS to be of a good balance. While not requiring much time or effort to be conducted overall, its questions have the potential to provide additional information about the systems that could be used to make recommendations. For instance, "I needed to learn a lot of things before I could get going with this system" and "I thought there was too much inconsistency in this system" both provide insight into some more specific strengths or weaknesses of the system in question.

The SUS is also a time-tested, widely used and scrutinised method. An empirical study of the SUS used nearly 10 years of data to conclude that the survey does effectively fulfill its purpose of attaining a user's subjective rating of a product's usability [5].

A study of SUS usage with non-native English speaking participants strongly recommended adjusting a single question to reduce confusion [13], and we chose to adopt this for our usability study. Finally, we modified the fourth question by changing 'technical person' to 'tutor/technical person', hopefully reducing potential confusion as the context in this usability study does not have an obvious definition of who a technical person is.

### 4.3 Study Execution

Students of the cloud computing class had access to a dedicated teaching cluster of 30 nodes for developing and testing their solutions. This cluster was shared among all students and had all systems installed using the same HDFS file system and YARN resource manager. The versions used were: Apache Hadoop MapReduce v2.7.2; Apache Spark v2.1.1; Apache Flink v1.2.1. Configuration changes from default settings were kept to a minimum. During the installation and testing process, to ensure that the systems would work from a the perspective of a logged-in student, we experienced cluster multitenancy issues and found little documentation to assist, ultimately requiring a large amount of our time to work around.

We aimed to support both Java and Python as implementation languages, and consequently lecture and tutorial materials were provided in both languages. However, despite our best efforts, we did not succeed in creating exercise materials for Flink with Python that worked well enough that we could instruct students on. We found that some simple operations were behaving unpredictably or producing overly complicated dataflows, making it very difficult to solve even basic tasks. There was limited documentation on the matter and the whole experience felt rather immature. Ultimately we did not provide teaching materials for Apache Flink in Python, and recommended that students avoid using it for their assignments. This recommendation was followed as all Apache Flink submissions were written in Java.

Due to the delays caused by this struggle, we were forced to release Apache Flink tutorial exercises (only in Java) one week after Apache Spark's, as opposed to the desired synchronised release. This remained prior to release of assignment 2, and lectures still covered both systems in parallel.

We designed the usability study such that the comparisons of Apache Flink and Apache Spark would be done in a crossed A/B test to avoid potential bias towards the first system used after MapReduce. To this end, student pairs were randomly assigned to use Apache Spark or Apache Flink for assignment 2, with an even split.

However, some students requested to change their assigned system for assignment 2 from Flink to Spark. We suspect that this was due to Flink's tutorial exercises having been delayed while working on Python support, meaning students had less experience with it than Apache Spark when assignment 2 was to come around. We did not deny their requests, and instead offered other Apache Spark groups to transfer to Apache Flink in an effort to restore balance. Ultimately there were more students using Spark than Flink for assignment 2; specifically 39 for Spark compared to 30 for Flink. All students cooperated in using the alternate system for assignment 3, which meant that assignment 3 had more Flink users.

Students were instructed not to use higher level APIs like FlinkML or Apache Spark's MLlib in their assignment solutions. This instruction was adhered to by all students.

Standard iterative algorithms relevant to big data contexts were selected to be the focus of the final two assignments. These algorithms were *k-means clustering* for assignment 2, and the *Apriori algorithm* for assignment 3. Example implementations of these algorithms are available for both Apache Spark and Apache Flink, however for different use cases and data sets. Despite this, we decided not to change the algorithms, as participants still would have to put a significant amount of effort into understanding and adapting those implementations to solve the assignment tasks. As expected, some students did find these existing implementations and did reference and adapt them to solve the assignments. Considering this, we decided not to include metrics such as their solutions' number of lines of code or other code related metadata in the usability study data set.

Our attempt to reduce bias in marking was not successful. Students made many mistakes that didn't fall beneath the predefined categories in the marking rubric, thus still requiring judgement of penalties. The end result was that two markers provided a significantly lower average mark than the other two; namely 10.5 compared to 13 where the maximum mark was 15. Considering this, it was decided that assignment marks should not be included in the usability study.

Despite the various mentioned setbacks, the usability study had a high participation level of over 80%, all students were cooperative with the crossed A/B test and in completing the surveys, and the end-of-semester course satisfaction feedback was high overall. The usability study had 72 participants, reduced to 69 after removing three with incomplete course participation.

Following completion of the semester, participants' three assignment surveys, and some metadata such as their degree code, was linked, recorded and anonymised. All feedback from surveys was checked and any personal information was removed.

### 4.3.1 Self-Reflection Surveys

Self-reflection surveys were available for individuals to complete immediately following the submission of each assignment. As the submission system provided no mechanism to enforce the surveys' completion, we instead reminded students to complete them during lectures and tutorials, and by direct email. Eventually all students completed the surveys, however the delay between assignment and survey completion is something we would've preferred to avoid. Approximately 60% of students submitted the self-reflection survey on the same day that they submitted the assignment.

The question about how much time was spent working on the assignment was updated following inspection of assignment 1 survey results, where we found that about half of the responses were 24+ hours. The question was updated to use thirteen 4 hour bins (instead of seven) in the assignment 2 and 3 surveys, from 0-4 hours to 44-48 hours and then 48+.

Despite all survey questions being optional, the four programming experience questions in the first survey, and all of the system preference questions, were answered by *all* students. Two out of 207 time questions were left unanswered, along with some SUS statement responses preventing calculation of 7 individual SUS scores out of 207 in total.

## 5 Analysis

### 5.1 Method

The self-reflection surveys created within the LMS were configured to *not* be fully anonymous, as if they were we would not be able to: correlate participant responses across multiple surveys; include other metadata such as the systems and programming languages used per assignment; or remove the data for students who opted not to participate in the study. Thus student identifiers were included with survey responses and would need to be removed manually at some point.

Data was initially exported from the LMS in Excel format following completion of the unit of study. The exported files contained many excess columns such as question IDs and labels, all of which were manually removed. The remaining columns contained only relevant metadata or answers to survey questions, and the column headers were renamed to reflect that, for instance from "Question 4 answer" to "Python Experience". Data for each of the three surveys were exported separately, creating multiple files which were joined to a single file using student identifiers. The file was then converted to a tab-separated values file, and rows for students who opted not to participate in the study were removed. Additional columns of data including participants' system and language used per assignment were added. Once all necessary data was removed, added, and double-checked, all student identifiers were removed. Then all textual feedback was scanned for any personal information for removal, of which none was found. From this point the survey data was entirely anonymised – the point of no return.

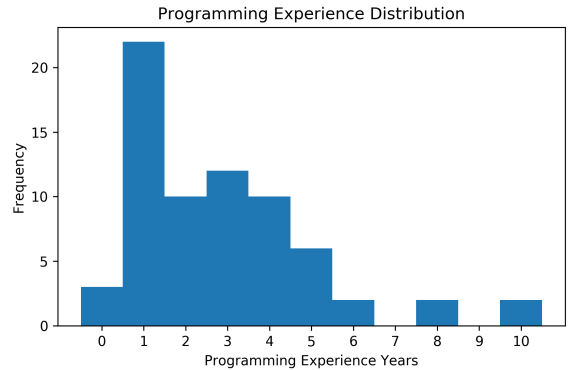

Figure 1: Frequency histogram displaying all 69 responses for the programming experience question. A value of 10 means 10 or more years.

With the data ready for analysis, it was then copied into a directory where a Python Jupyter Notebook would be created and executed. The notebook is a common choice for exploratory data analysis and among data scientists, and thus seemed appropriate for the task of analysing the usability study data. The following Python modules and libraries were used in performing the exploration and analysis: NumPy, a "fundamental package for scientific computing with Python"; Matplotlib for creation of charts and plots to be both displayed in-line in the notebook and saved to files; pandas to simplify data usage and manipulation; and SciPy for its library of statistical functions.

Exclusion of Python user data from the usability study, as for Section 5.7, was performed using a single conditional statement near the beginning of the notebook. This statement could easily be toggled (by changing `True` to `False`), followed by a restart and full run of the notebook, to quickly compare analyses with and without including Python data.

## 5.2 Background of Participants

Of the 69 participants: 55 were graduate computing students; 7 were Master of Data Science students, who do not necessarily have a computer science background; 6 were final-year undergraduate students; and 1 was a master's student of a different degree.

Figure 1 shows that most (78.3%) participants reported having 1 to 4 years of programming experience. Reflecting the diversity of the student cohort, around half of the participants reported to have limited or no proficiency in shell and Python environments, compared to around a third for Java, as seen in Figure 2. There is a slightly negative correlation between Java and Python proficiency, with a Spearman's rank correlation coefficient of -0.128. The Pearson correlation coefficient was not used because the data is ordinal level, failing the test's ratio level assumption.

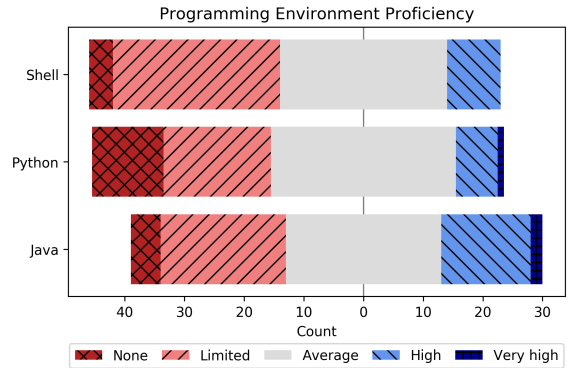

Figure 2: Diverging stacked bar chart [17] displaying all 69 Likert scale responses for questions on perceived programming proficiency.

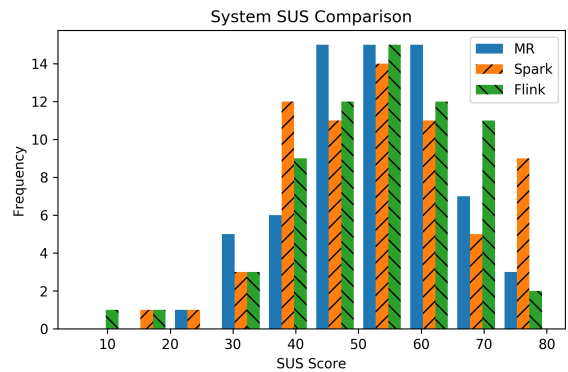

Figure 3: Frequency histogram comparing all 69 SUS scores per system (minus 7 individual incomplete responses).

### 5.3 Preferences and SUS Scores

Following the completion of the third survey, participants reported their system preference as: 8 (11.6%) for Apache Hadoop MapReduce; 29 (42.0%) for Apache Spark; and 32 (46.4%) for Apache Flink. This is strong evidence that Spark

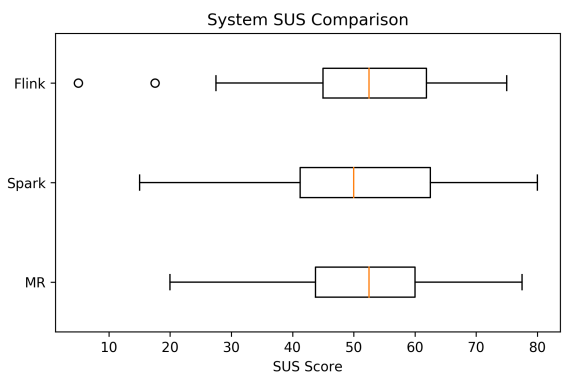

Figure 4: Box and whisker plot comparing all 69 SUS scores per system (minus 7 individual incomplete responses). Whiskers extend 1.5 IQR beyond quartiles. Outliers are circles.

and Flink were preferred over Hadoop MapReduce.

The difference was similarly pronounced among data science students where 5 of 7 preferred Flink over Spark or MapReduce. However we note that four of those five students did use Flink in assignment 2 before using Spark, which could have an influence, as Section 5.4 will show. Due to the small amount of data in this context there was no applicable significance test.

The SUS scores reveal little information, with all systems sharing similar distributions and quartiles, as visible in Figures 3 and 4. This can be supported by a Friedman test of all participants' three system SUS scores, resulting in a probability ($p$) value of 0.943, which suggests that there is no statistically significant difference between the systems. One-way ANOVA of repeated measures was not used because the data was nonparametric. More specifically, the hypothesis that Apache Flink scores come from a population with a normal distribution can be rejected with a Shapiro-Wilk $p$-value of 0.039 at a significance level of 0.05, and also have outliers as visible in Figure 4. (Hadoop MapReduce and Apache Spark have Shapiro-Wilk $p$-values of 0.767 and 0.352 respectively.)

While participants have strongly suggested preference of Spark or Flink over MapReduce, there is no clear distinction between the two data processing systems themselves. It also means that the SUS, though a standard measure for system usability, appears to poorly correlate with perceived preferences in this context, as its lack of difference between the systems does not at all reflect the strong separation of MapReduce.

## 5.4 Influence of Assignments

The usage of a crossed A/B test in assignments 2 and 3 means the SUS scores per assignment differ from the SUS scores per system. The difference in SUS scores is more clearly pronounced per assignment than per system, as visible by comparing Figure 5 to Figure 4. It appears as though assignment 2 had the highest relative SUS scores, and assignment 3 the lowest. This is supported by a Friedman test of all participants' three assignment SUS scores, resulting in a $p$-value of 0.025, which suggests a statistically significant difference at a significance level of 0.05.

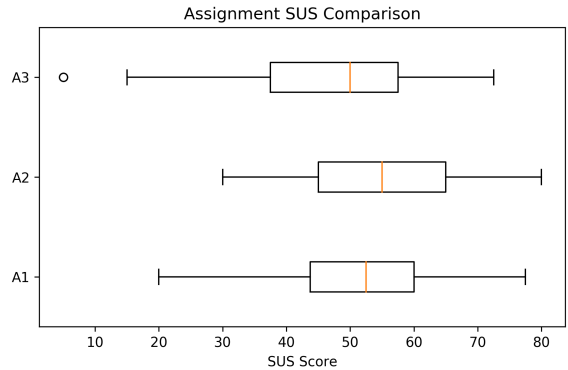

Figure 5: Box and whisker plot comparing all 69 SUS scores per assignment (minus 7 individual incomplete responses). Whiskers extend 1.5 IQR beyond quartiles. Outliers are circles.

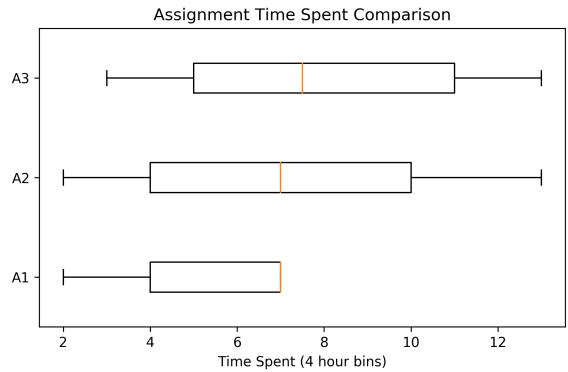

Figure 6: Box and whisker plot comparing all 69 amounts of time spent per assignment (minus 2 individual incomplete responses). Whiskers extend 1.5 IQR beyond quartiles. A value of 3 means 8-12 hours, and 13 means 48+ hours. Assignment 1 was limited to value 7 or 24+ hours.

More than half of the participants (39 participants or 56.5%) preferred the system they used in assignment 2, compared to 22 (31.9%) for assignment 3 (with the other 8 (11.6%) preferring Apache Hadoop MapReduce), which is quite a noteworthy difference. However, it's difficult to reason about this difference, as there's no clear distinction as to whether it's due to some form of a first-system-used bias or differences in assignment difficulty (as described in Section 4). Figure 6 shows the time spent working on assignments, which provides a hint as to potential differences in assignment difficulty, wherein assignment 3 appeared to require slightly more time than assignment 2. However, this claim is not supported by a one-sided sign test with plus representing participants who spent more time on assignment 3 than 2, and minus otherwise, resulting in a $p$-value of 0.358. The sign test is used because the data is non-normal and asymmetrical, as clearly visible in Figure 7, ruling out the paired t-test due to it being a parametric test, and the Wilcoxon signed-rank test due to it having high Type I error rates when used with asymmetric data.

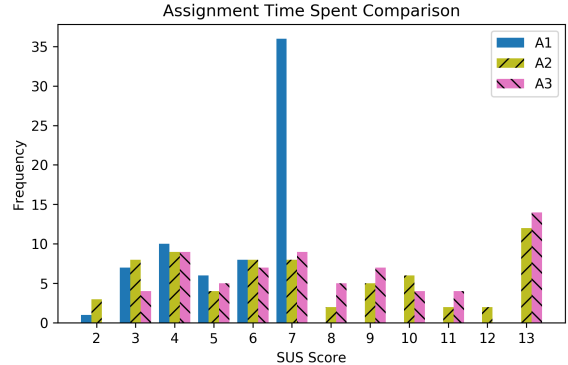

Figure 7: Frequency histogram comparing all 69 amounts of time spent per assignment (minus 2 individual incomplete responses). A value of 3 means 8-12 hours, and 13 means 48+ hours. Assignment 1 was limited to value 7 or 24+ hours.

While we suspect assignment difficulty and 'first-used advantages' could have affected perceived preferences, we've not been able to quantitatively explain the significant difference between assignment SUS scores, nor any link between SUS scores and assignment preferences. With that being said, the crossed A/B test that was used should have helped to reduce any effect of these biases on the systems themselves.

## 5.5 Programming Duration versus System

Apache Spark and Apache Flink shared similar reported development times for assignment 2, but with Flink perhaps showing strength for assignment 3, which you can see in Figure 8. However, this is not supported by Mood's median tests of the two systems' (independent) time spent data, resulting in $p$-values of 0.661 for assignment 2 and 0.624 for assignment 3, and thus suggesting no statistically significant difference in either. Mood's median test is used because the data is non-normal and differs in distribution, as clearly visible in Figure 9, ruling out the one-way ANOVA due to it being a parametric test, and the Mann-Whitney

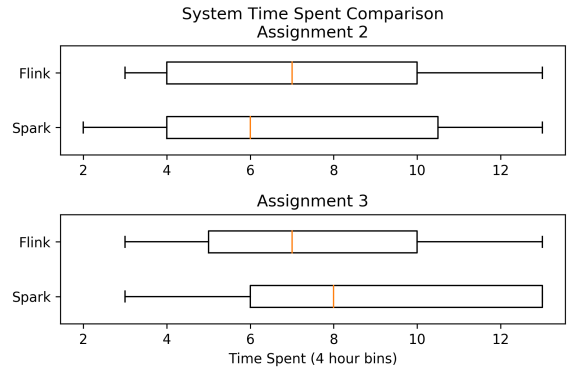

Figure 8: Box and whisker plots for assignments 2 and 3 comparing all 69 amounts of time spent per assignment (minus 1 individual incomplete response). Whiskers extend 1.5 IQR beyond quartiles. A value of 3 means 8-12 hours, and 13 means 48+ hours.

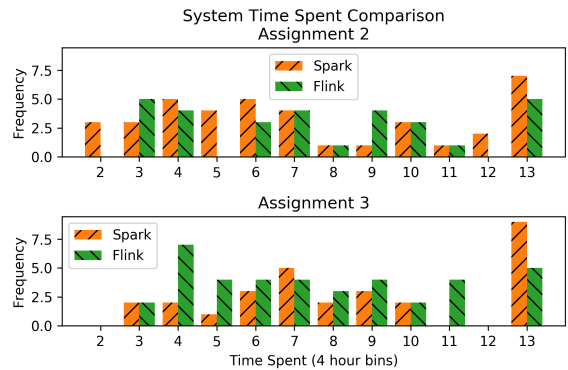

Figure 9: Frequency histograms for assignments 2 and 3 comparing all 69 amounts of time spent per system (minus 1 individual incomplete response). A value of 3 means 8-12 hours, and 13 means 48+ hours.

U test due to it not testing changes in medians or means (but instead testing changes in distribution) when used with data of differing distributions.

Spark and Flink do not present a significant difference in the amount of time that was required to complete either of the assignments. This shows that both systems are similarly suitable for completion of data analysis tasks like those in assignments 2 and 3.

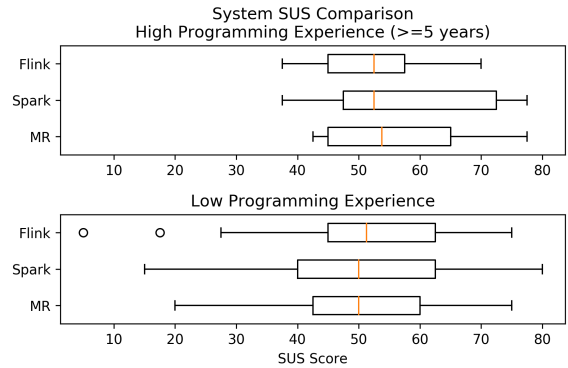

Figure 10: Box and whisker plots for high (>=5 years) and low programming experience participants comparing all 69 SUS scores per system (minus 7 individual incomplete responses). Whiskers extend 1.5 IQR beyond quartiles. Outliers are circles.

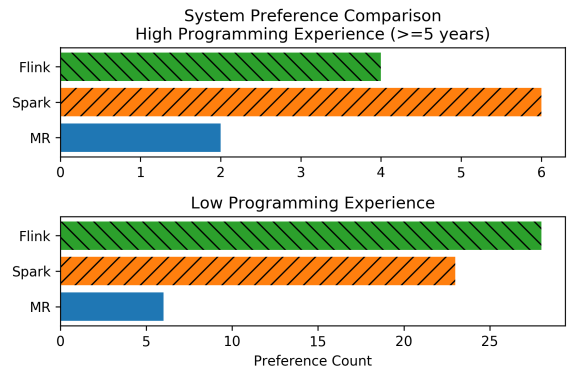

Figure 11: Bar charts for high (>=5 years) and low programming experience participants comparing reported system preference counts upon completion of assignment 3. There are 69 preferences measured in total. Charts have differing Preference Count scales.

## 5.6 Influence of Programming Experience

As previously mentioned, Figure 1 shows that most (78.3%) participants reported having 1 to 4 years of programming experience. In this subsection we'll consider the 12 (17.4%) participants who reported having more than 4 years experience as being of relatively 'high experience'. Their experience is distributed as so: 6 with 5 years experience, and 2 with each of 6, 8 and 10+ years experience. Is there any difference in the reported preferences or SUS scores for high experience participants compared to the majority?

A comparison between the two groups' system SUS scores is shown in Fig-ure 10, which does not tell much of any preference or skew. The ranges of all three systems are more dense and perhaps slightly higher overall in the high experience group, and also particularly lacking the long tail towards lower scores. However, it would be difficult to indicate that it is not due to coincidence. These observations still apply, perhaps with a slight reduction in density, when moving the threshold from 5 years to 3 years and thus placing 34 (49.3%) participants in the high experience tier.

A comparison between the two groups' reported preferences is shown in Figure 11, which also does not tell much of any preference or skew. Hadoop MapReduce remains behind the dataflow engines, but it is difficult to reason much further. For instance, the difference in Apache Flink and Apache Spark looks quite pronounced, with Flink falling behind in comparison to the low experience group. However, in actual fact it is only a difference of two individual preferences, which would be difficult to reason as anything other than coincidence. When moving the threshold from 5 years to 3 years and thus placing 34 (49.3%) participants in the high experience tier, the two charts end up look nearly identical, also with very similar $x$ scales.

It was good to have collected the programming experience data as part of the survey, especially with full participation in that regard, having provided information regarding the background of participants and the diversity present in the study. However, this information did not turn out to be indicative of any pattern in reported preferences or system SUS scores.

## 5.7 Influence of Programming Language

In the analysis thus far, participants' usage of Python or Java for each system or assignment has not been taken into account. However, it is mentioned in Section 4.3 that learning materials, or more specifically tutorial exercises and sample assignment solutions, were not successfully created for Apache Flink; that students were recommended not to use Python for Flink; and that this recommendation was followed by all students. Unfortunately, this presents an inconsistency in the experiment which could bias some of the findings. This subsection will explore said biases by comparing the previous analyses with what would be the case had all Python user data been excluded.

Of the 69 participants whose data was used in the above analyses, 32 (46.4%) had used Python for either or both of assignment 1 (Hadoop MapReduce) or the assignment in which they used Apache Spark. Excluding this data leaves 37 records to work with. Furthermore, the crossed A/B test that was employed did not take programming language usage into account, and thus the Python records removed could have further produced skew in that regard. With 69 records the Spark vs. Flink usage split for assignment 2 (and conversely for assignment 3) is 39 vs. 30 (56.5% vs. 43.5%) respectively, skewed due to students' requests to switch to Apache Spark as described in Section 4.3, whereas with 37 records it is 23 vs. 14 (62.2% vs. 37.8%) respectively. Flink is 9 users behind Spark (for assignment 2) in both cases, however this difference is more pronounced as a percentage in the latter case.

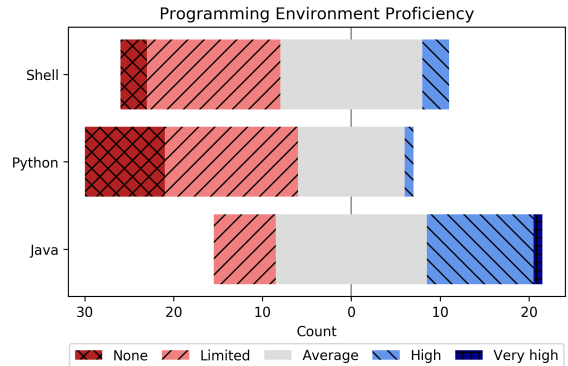

Figure 12: Diverging stacked bar chart [17] displaying all 37 (Python user data excluded) Likert scale responses for questions on perceived programming proficiency.

Of the subsections that were analysed earlier, noteworthy differences were only apparent in relation to the background of participants, and their reported system preferences.

### 5.7.1 Background of Participants

Of the 37 participants: 33 were graduate computing students; 1 (compared to 7) was a Master of Data Science student; 2 were final-year undergraduate students; and 1 was a master's student of a different degree. Furthermore, Figure 12 (compared to Figure 2) shows the dominance of Java among the cohort, also with substantially more participants reporting having limited or no Python proficiency than average or higher.

This apparent reduction in diversity may be due to individuals in this cohort having been more likely to have come from a computer science background, where compiled and object-oriented programming languages would be of a larger focus than scripting languages. Unfortunately, this representation may not effectively capture the experiences of interdisciplinary users, as has been one of our intentions in conducting this study.

### 5.7.2 Preferences

Following completion of the third survey, participants reported their system preference as: 4 (10.8%) for Hadoop MapReduce; 12 (32.4%) for Apache Spark; and 21 (56.8%) for Apache Flink. While this result is similar to the full (Python included) data set in how MapReduce is behind in comparison to the dataflow engines, it differs from having Spark and Flink on a very close footing to instead placing Flink quite noticeably ahead as the participants' preferred system.

This difference seems to make sense considering that Python users would likely have felt inconvenienced by the inability to use Python for Flink in com-parison to Spark or MapReduce, thus reducing Flink's popularity among the full cohort. With Python users excluded from analysis, it has now become apparent that Flink is preferred over Spark among Java users.

Unfortunately, as described in Section 4.3.1, there is little information available from participants to justify these reported preferences, and so we can make no conclusion as to why Java users may have preferred Flink over Spark, especially considering the similarities in their usage as described in Section 3. The key difference in usage that we could think of was in Flink's availability of `IterativeDataSet`, which could have been used in solving both assignments 2 and 3.

With the full data set it was apparent that participants tended to prefer the system they used for assignment 2 more than assignment 3, as described in Section 5.4. With Python data excluded, this changes to having 17 (45.9%) participants preferring the system used for assignment 2 compared to 16 (43.2%) for assignment 3 (with the other 4 (10.8%) preferring MapReduce). It makes sense that assignment 3 would've gained preference considering the skewed crossed A/B test distribution, where more participants used Flink for assignment 3, in combination with participants' higher preference of Flink. This means that although reported preferences put assignments 2 and 3 on similar proportions, it remains likely that assignment 3 was somehow disadvantaged.

## 5.8 Individual SUS Statements

This subsection will discuss Figure 13, which displays all participants' responses to all ten SUS statements. The figure has been organised in a particular way considering the nature of the SUS. It's caption describes it's intended interpretation.

The first takeaway is that, for all ten questions, there is little difference in response distribution between the three systems. This explains why the SUS scores were very similar, as discussed in Section 5.3. It shows that in terms of usability as examined by the System Usability Scale, these three systems apparently share the same strengths and weaknesses, which is unintuitive considering that Apache Spark and Apache Flink feature much higher level APIs.

Furthermore, most questions had a good response as either a slight or strong majority – a good sign for the usability of these systems. There are three exceptions: "Q4: I think that I would need the support of a tutor/technical person to be able to use this system", which most participants agreed with; "Q7: I would imagine that most people would learn to use this system very quickly", which a slight majority of participants disagreed with; and "Q10: I needed to learn a lot of things before I could get going with this system", which participants overwhelmingly agreed or strongly agreed with.

These three questions with bad responses cover the topics of learning and support, and are the only questions in the set of ten concerning them. The other seven questions with good responses covered topics such as ease of use and system design and complexity. Is it normal that participants perceived the systems to be well designed, non-complex and usable, and yet hard to learn and in need of support to be used? One interpretation of this result is that participants felt that the systems were to be complex by design, considering their function as distributed computing engines, and in that context perceived them to be usable and such.

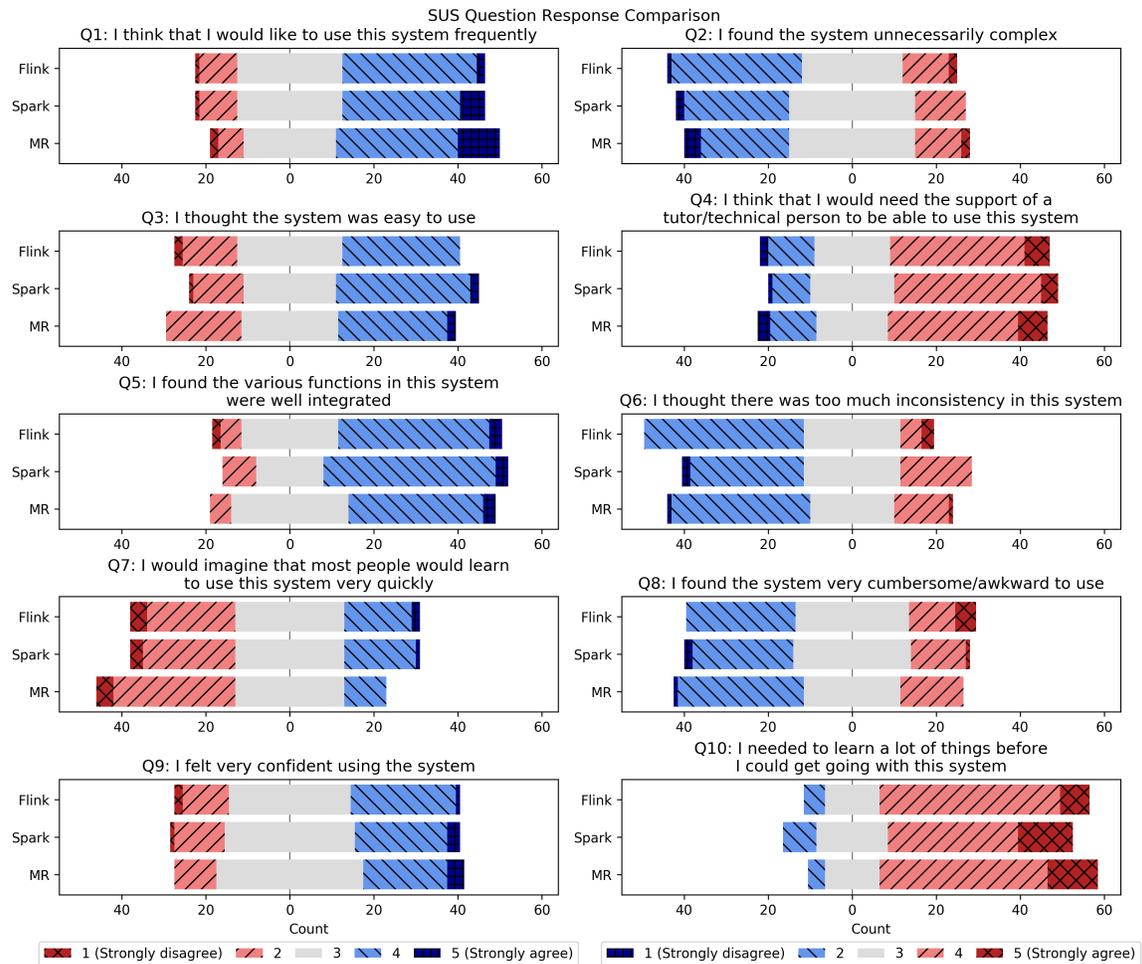

Figure 13: Diverging stacked bar charts [17] displaying all 69 Likert scale responses for each SUS statement, however minus some missing individual responses. The five odd numbered statements on the left are 'positively natured', such as "Q3: I thought the system was easy to use" – agreeing with this statement is *good* in terms of the system's usability. The five even numbered statements on the right are 'negatively natured', such as "Q2: I found the system unnecessarily complex" – agreeing with this statement is *bad* in terms of the system's usability. To display this effectively, the legends on the left and right have their colours inverted – blue is used to highlight good responses (regardless of a statements' positive/negative nature) and red for bad responses.

However, the steep learning curve still needed to be addressed for them to use the system. We believe this is indicative of a prominent barrier to adoption of data science systems – although they effectively abstract various distributed computing and dataflow challenges away, without simplifying the process of learning to use and overcoming difficulties with these systems, users with non-computing backgrounds will have significant trouble adopting them. From our experience and from the feedback of participants, we found the key areas for improvement to be first-party documentation and the process of debugging programs.

# 6 Conclusion

We performed a usability study with students in a cloud computing class, to address the lack of usability data concerning modern distributed data processing platforms. The systems compared were Apache Hadoop MapReduce [3], and the more dataflow oriented Apache Spark [4, 30] and Apache Flink [2, 9]. The usability study primarily involved survey data collected from three surveys – one following completion of each of three data analytics assignments. The first assignment used Hadoop MapReduce, and the latter two employed a crossed A/B test with Spark and Flink.

The experiment worked well: study participation was high; students cooperated in the crossed A/B test without friction; only a small portion of survey data was left unfilled; and student course satisfaction levels remained high. Catering for the diversity of a class with both IT and data science students is a challenge, and we see the learnings and careful design of our study as one of our contributions. We do recommend setting aside a generous amount of time for preparation before the class begins.

We found that participants' perceived preferences were strongly in favour of either Spark or Flink in comparison to MapReduce, however there was little difference between the two modern systems themselves. There was also no significant difference in the amount of time participants reported they required to complete the assignments using either of the modern systems. Thus from a usability point of view, both Spark and Flink seem to be equally suitable choices over MapReduce, most likely due to the high-level nature of these data processing platforms.

We experimented with using the System Usability Scale (SUS) [8] to measure and compare the usability of the three systems, which we have not seen used in programming contexts despite being found effective elsewhere [5]. However despite its convenience, we ultimately did not find it highly effective, as it did not provide much insight into the usability of each system, nor did it correlate with perceived preferences. Looking in detail at responses to individual SUS questions highlighted weaknesses in the learn-ability and need for support in all three systems, supporting what we believe to be key areas for improvement in data science systems which would greatly support their adoption by users with non-computing backgrounds: first-party documentation and the process of debugging programs.

We gained additional insights from the participants' free-text feedback, and the experiences of instructors throughout the study, which highlight several areas of potential future improvement of the evaluated systems:

- Debugging in MapReduce was particularly difficult.

- MapReduce code tended to be overly verbose.

- Flink development environment setup was troublesome.

- Python-support in Flink 1.2 felt immature.

- Spark and Flink were quite similar to work with.

- Spark and Flink documentation covered basic usage quite well, but was limited for non-standard operations. Consequently, both Spark and Flink involved significant trial and error.

- Spark community support was good, but first-party documentation was lacking. Flink was described inversely.

A further finding was that participants preferred the first of the data processing engines that they encountered in the class – either Spark or Flink. Interestingly, there was also a significant difference in SUS scores between the assignments, however there was no suitable data to highlight the cause of this difference. We suspect that it's a combination of a first-used advantage and differences in assignment difficulty.

However, Python with Flink was not working to an acceptable degree and we thus could not provide support and learning resources for them. We recognise that this would have introduced some bias to the study, and to account for this analyses were repeated with Python user data excluded, where it became apparent that among Java users Flink was more often reported as preferred than Apache Spark.

Overall this experiment indicates that there are meaningful differences between big data processing systems in their usability, which might contribute to their adoption as much as technical aspects or raw performance – especially as all of them scale well when adding more virtual machines. We found that many data science users have a non-traditional computing background, and that consequently the focus needs to be more on usability factors such as ease-of-use, learnability, language support, auto-configuration, and community support. This study is the first step by our group to better understand the usability of big data processing systems for data science. Our long-term goal is to identify factors and to develop techniques for improving the usability – and hence the impact – of the next generation of big data systems.

**Acknowledgements**

Ethics approval was attained prior to the commencement of this study from the University of Sydney's Human Research Ethics Committee under project number 2017/212. This application was supported by the university's Research Cluster for Human-centred Technology.